\documentclass[journal]{IEEEtran}
\usepackage{amsmath}
\usepackage{dblfloatfix} 
\usepackage{booktabs}
\usepackage{multirow}
\usepackage{rotating}
\usepackage{url}
\usepackage{balance}
\usepackage{mathtools}
\usepackage{graphicx}
\usepackage{xcolor}
\usepackage{color}
\usepackage{soul}
\usepackage{placeins}
\usepackage{float}
\usepackage{hyperref}
\usepackage{amssymb}
\usepackage{pifont}
\usepackage{tabularx,colortbl}
\usepackage{amssymb}
\usepackage{bm}
\usepackage{tabularx}
\usepackage{threeparttable}
\usepackage{tablefootnote}
\usepackage{amsthm}
\usepackage[caption=false]{subfig}

\begin{document}
%
\title{Cascaded Multiwire-PLC/Multiple-VLC System: Characterization and Performance}

\author{
Hugerles S. Silva, \textit{Senior Member, IEEE}, Higo T. P. Silva, Paulo V. B. Tomé, Felipe A. P. Figueiredo, Edson P. da Silva, \textit{Senior Member, IEEE}, and Rausley A. A. de Souza, \textit{Senior Member, IEEE.}

\thanks{

H. S. Silva, H. T. P. Silva and P. V. B. Tomé are with University of Brasília, Federal District, Brazil (e-mail: hugerles.silva@unb.br, 241131464@aluno.unb.br, higo.silva@unb.br).

E. P. da Silva is with the Federal University of Campina Grande (UFCG), Paraíba, Brazil (edson.silva@dee.ufcg.edu.br).

F. A. P. de Figueiredo and R. A. A. de Souza are with the National Institute of Telecommunications (Inatel), Santa Rita do Sapucaí, MG 37536-001, Brazil (felipe.figueiredo@inatel.br, rausley@inatel.br).
}}

\markboth{Submitted to IEEE LATINCOM 2026}{}%
\maketitle

\begin{abstract}
This paper presents  a cascaded multiwire-power line communication (PLC)/multiple-visible light communication (VLC) system.
This hybrid architecture offers low installation cost, enhanced performance, practical feasibility, and a wide range of applications.
Novel analytical expressions are derived for key statistics and outage probability and bit error probability metrics. 
These analytical findings are corroborated via Monte Carlo simulations, with performance curves evaluated under diverse channel conditions and PLC/VLC system parameters.
All expressions developed in this work are original and have not appeared in previous literature.
Our proposed system proves feasible for smart environments, green communication systems, internet of things networks, industrial environments, and next-generation networks.

\end{abstract}
\begin{IEEEkeywords}
Average bit error probability, outage probability, relaying, PLC, VLC.
\end{IEEEkeywords}

\IEEEpeerreviewmaketitle
\section{Introduction}\label{sec1}

\IEEEPARstart{R}{ecent} advancements in power line communication~(PLC) and visible light communication~(VLC) are drawing interest from academia and industry due to their advantages.
In general, PLC technology is more secure than traditional wireless communication systems and offers cost-effective deployment using wired electrical network infrastructure to transmit data~\cite{Majumder}. 
In turn, VLC leverages light-emitting diodes (LEDs) for wireless transmission, offering high data rates, low power consumption, and immunity to electromagnetic interference, which makes it particularly suitable for environments where radio frequency (RF) communication is limited or prohibited~\cite{Komine2004}.

Recent works present performance evaluations of PLC and VLC systems under different operation modes.
A capacity analysis of PLC over Rayleigh fading channels, subject to colored Nakagami-$m$ additive noise, is presented in~\cite{Ai2016}. 
Expressions for the probability density function~(PDF) and cumulative distribution function~(CDF) of the instantaneous signal-to-noise ratio~(SNR) are derived. 
Based on the mentioned first-order statistics, metrics are presented to investigate the impact of the channel characteristics on the PLC system capacity. 
The cascaded multiwire-PLC/multiple-input multiple-output (MIMO)-RF communication system is analyzed in~\cite{Ai2021}, proposing an opportunistic decode-and-forward relaying system that combines multiwire-PLC and RF technologies to leverage their respective advantages. 
The mentioned study derives closed-form expressions for outage probability~(OP), average bit error probability~(BEP), and average channel capacity, which are validated through Monte-Carlo simulations. 

In turn, the performance of non-orthogonal multiple access~(NOMA) under a downlink VLC system is analyzed in~\cite{Yin2016}.
The authors deduce analytical expressions for the system coverage probability and ergodic sum rate, and simulations corroborate the theoretical framework. 
The cascaded free-space optical~(FSO)-VLC communication system, proposed for indoor multimedia broadcasting, is investigated in~\cite{Gupta2017}. 
The authors derive expressions for the PDF and CDF of the end-to-end SNR and the OP and BEP under different channel and system parameters. 
In~\cite{Vats2018}, an analysis of error and outage probabilities in a three-hop hybrid VLC-FSO-VLC-based relayed optical wireless communication system is presented. 
This study examines the effects of atmospheric turbulence, pointing errors, the semi-angle, and the detector's field of view. 
The aforementioned study considers the VLC and FSO links modeled using the Lambertian emission model and Gamma-Gamma fading statistics, respectively. 
Closed-form expressions for OP, average symbol error probability, and asymptotic OP are derived and validated through numerical simulations.

The combination of properties and joint advantages of PLC with VLC is also evaluated.
In~\cite{Ndjiongue}, a hybrid PLC-VLC system is proposed, employing quadrature phase shift keying combined with orthogonal frequency division multiplexing for the PLC channel, while color shift keying is utilized over the VLC channel to transmit information.
The system performance is evaluated through simulated bit error rate (BER) curves.
In~\cite{Jani2019}, the performance analysis of a cooperative PLC-VLC system with multiple access points for indoor broadcasting is investigated using OP and BER.
Extending~\cite{Jani2019}, the performance analysis of a mixed cooperative PLC-VLC system for indoor communication is presented in~\cite{Jani2020}, incorporating the effects of lognormal fading, additive background noise, impulsive noise, and user positions on the VLC link.
Analytical closed-form expressions for the CDF and PDF of the end-to-end SNR, as well as expressions for the OP and average BER under different system and channel parameters, are derived. 
The performance of a cooperative PLC-VLC indoor broadcasting system is investigated in~\cite{Jani2021}, which consists of mobile user nodes for Internet of Things (IoT) networks.
The system integrates mobile end-user nodes, with the PLC link serving as the backbone of the VLC link, which is connected via a DF relay. 
Novel closed-form expressions for the PDF and CDF of the equivalent end-to-end SNR are presented alongside OP and average BEP results, which have been validated through simulations. 
A comparison among related works and our paper is presented in Table~\ref{tab:comparison}. 

This paper introduces the cascaded multiwire-PLC/multiple-VLC system to leverage the advantages of both techniques. 
This hybrid architecture offers low installation costs, enhanced performance, and practical feasibility. 
We derive closed-form expressions for key statistical measures and performance metrics. 
To the best of our knowledge, the analyses and all the expressions presented here are novel in the literature. 

In summary, the main contributions of this article are: (i) A novel system that combines multiwire-PLC, DF relay, and multiple-VLC.
 (ii) New and exact expressions for the PDF and CDF of the equivalent end-to-end SNR in the proposed cascaded multiwire-PLC/multiple-VLC system.
(iii) Expressions for the OP and average BEP.

The rest of the paper is structured as follows.
Section~\ref{SystemModel} presents the system and channel modeling.
The equivalent end-to-end SNR statistics for the cascaded multiwire-PLC/multiple-VLC system are derived in Section~\ref{Statistics}. 
Some important performance metrics\textemdash, specifically, the exact OP, BEP, and channel capacity expressions\textemdash are presented in Section~\ref{PerformanceAnalysis}.
Section~\ref{results} shows the numerical results.
Section~\ref{conclusao} brings the conclusions.

\begin{table}
\centering
\caption{Comparison Among Related Works.}
\renewcommand{\arraystretch}{1.3}
\begin{tabular}{l|cccc|c}
\toprule
\textbf{Work} & \cite{Ndjiongue} & \cite{Jani2019} & \cite{Jani2020} & \cite{Jani2021} & \textbf{Our Paper} \\\hline
\midrule
{Multiwire-PLC} &  &  &  &  & \checkmark \\\hline
{Multiple-VLC} &  & \checkmark &  &  & \checkmark \\\hline
{Relay Protocol}            &  & \checkmark & \checkmark & \checkmark & \checkmark \\\hline
{Statistics of the system}&  & \checkmark & \checkmark & \checkmark & \checkmark \\\hline
OP               &  & \checkmark & \checkmark & \checkmark & \checkmark \\\hline
BEP               & \checkmark & \checkmark & \checkmark & \checkmark & \checkmark \\
\bottomrule
\end{tabular}
\label{tab:comparison}
\end{table}

\section{System and Channel Models}\label{SystemModel}

\subsection{System Model}

\begin{figure}
    \centering
    \includegraphics[width=0.8\linewidth]{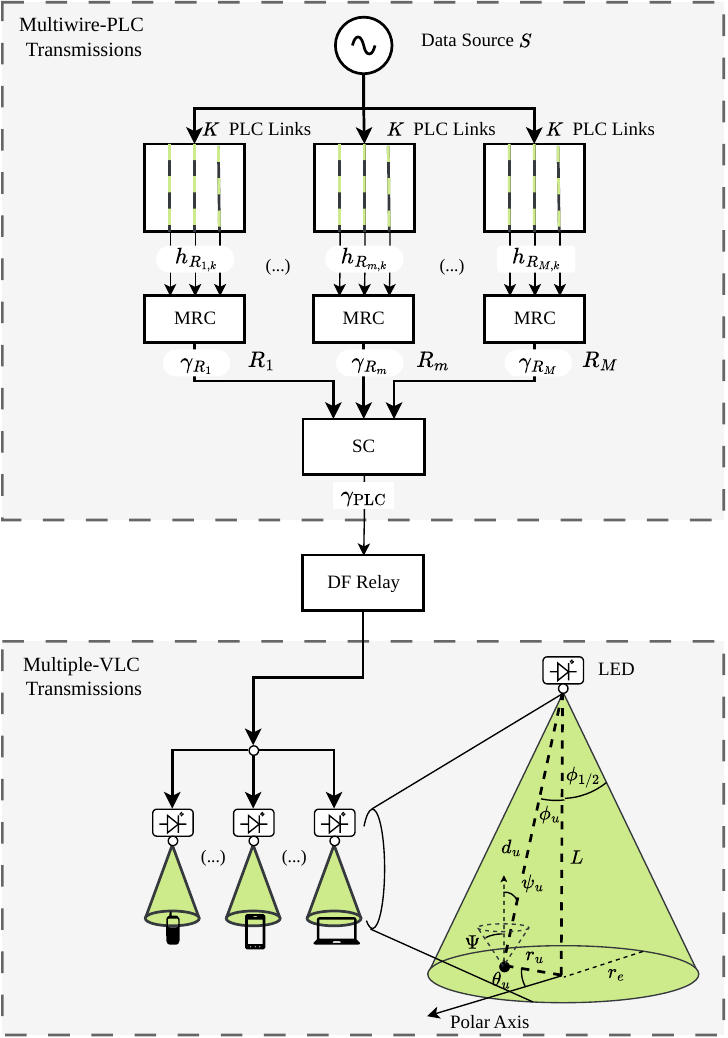}
    \caption{Cascaded multiwire-PLC/multiple-VLC system model.}
    \label{fig:cascadedSystem}
\end{figure}

Fig.~\ref{fig:cascadedSystem} illustrates the adopted cascaded multiwire-PLC/multiple-VLC system indoor system architecture. 
The communication process operates in two distinct phases. 
Initially, the PLC link acts as a backhaul connection, establishing access to the backbone network. 
In the second phase, the data is transmitted to the end users through multiple VLC access points positioned on the ceiling of the indoor environment. 
It is assumed that mobile users are uniformly distributed within the coverage area. 
The indoor VLC system is integrated with the PLC link through a decode-and-forward (DF) relay, ensuring efficient signal relay and data continuity. 
The described configuration facilitates data transmission between a source node and a specified receiver within the illumination region provided by the LED. 

In the PLC subsystem of the cascaded model, data originating from the source node, denoted as $S$, is relayed to $M$ intermediate nodes, represented as $R_m$, with $m = 1, 2, \dots, M$. 
Each relay node utilizes $K$-branch receiver diversity to counteract the negative impacts commonly associated with PLC channels~\cite{Ai2021}, where $K$ represents the number of PLC links per node. 
The diversity gain is achieved through multiple physical power lines, including neutral, live, and ground wires, as well as potential alternative paths. 
It should be noted that the PLC channel gain amplitudes are modeled as lognormally distributed random variables (RVs), following the assumptions presented in~\cite{Guzelgoz}, to reflect realistic channel behavior.

In general, the channel characteristics of a VLC system are influenced by both line-of-sight (LoS) and diffused non-line-of-sight (NLoS) components generated by dispersive interactions with the surfaces within the indoor environment. 
Considering the high losses involved in the diffuse light scattering mechanisms, the LoS component contributes significantly more to the link than the NLoS components~\cite{Farshad}. 
Given this dominance, this study's analysis and performance evaluation of the VLC link consider only the LoS component, thereby simplifying the modeling while retaining accuracy.

\subsection{Cascaded Multiwire-PLC Channel}

The signal received at the $m$-th  intermediate relay node, impacted by the $k$-th propagation branch of the PLC channel, with $k = \{1,\dots,K\}$, is represented by $y_{R_{m,k}}$.
Based on~\cite[Eq. (1)]{Ai2021}, this received signal can be  expressed as
    \begin{equation}
    \label{eq:rec_signal_mk}
    y_{R_{m,k}} = \sqrt{P_{\text{PLC}}\beta_{\text{PLC}}} h_{R_{m,k}} s + n_{R_{m,k}}, \end{equation}
where $s$ is the transmitted signal, which is normalized to have unit power, and $P_{\text{PLC}}$ represents the total transmitted power. 
Furthermore, $\beta_{\text{PLC}} = \exp(-2(\alpha_1+\alpha_2f^{k})\ell_{\text{PLC}})$ is the attenuation along the PLC system cables, with constants $\alpha_1$ and $\alpha_2$, frequency $f$, expressed in MHz, and powerline link length $\ell_{\text{PLC}}$.
The term $h_{R_{m,k}}$ denotes the channel gain corresponding to the $k$-th communication path and the $m$-th node. 
Finally, the component $n_{R_{m,k}}$ in~\eqref{eq:rec_signal_mk} represents the noise components, which can account for distortions, external interference, thermal noise, impulsive noise, and others.

The channel gains $h_{R_{m,k}}$ are assumed to be independent and identically distributed (i.i.d.) RVs that follow a lognormal distribution \cite{Ai2021} whose PDF is given by
    \begin{equation}\label{eq:PDFLOGNORMAL} 
    f_{h_{R_{m,k}}}(x) = \frac{1}{\sqrt{2 \pi \sigma_{h}^2} x} \exp \left[ -\frac{(\ln x - \mu_{h})^2}{2 \sigma_{h}^2} \right], 
    \end{equation}
in which $\mu_{h}$ and $\sigma_{h}^2$ are the logarithmic mean and variance of $h_{R_{m,k}}$, respectively. 
Without loss of generality, the statistics $\mu_{h}$ and $\sigma_{h}^2$ are set so that $\mathbb{E}\{h_{R_{m,k}}^2\} = 1/K$, in which $\mathbb{E}\{\cdot\}$ is the expected value operator.
The lognormal distribution accurately models the multiplicative fading effects typically observed in PLC environments, capturing the impact of attenuation, impedance mismatches, and other channel impairments.

The noise present in a PLC channel consists of two primary components: background additive white Gaussian noise (AWGN) and a zero-mean impulsive noise component that follows a Gaussian distribution. 
Impulsive noise is caused by sudden bursts of interference caused by electrical appliances, switching events, or other transient disturbances in the power grid. 
As a result, the overall PLC channel noise is modeled as a Gaussian RV with a total variance given by \cite{Ai2021}
    \begin{equation}
    \sigma_{\text{PLC}}^2 = (1 - p) \sigma_\text{g}^2 + p (\sigma_\text{g}^2 + \sigma_\text{i}^2),
    \end{equation}  
in which \( \sigma_\text{g}^2 \) represents the background AWGN power, while \( \sigma_\text{i}^2 \) corresponds to the power of the impulsive noise component. 
The parameter \(p\) denotes the probability of impulsive noise occurrence, capturing the intermittent nature of such disturbances in PLC environments. 
This formulation accounts for the fact that with probability \( 1 - p \), the noise consists solely of AWGN. 
In contrast, with probability \( p \), the impulsive noise component is also present, increasing the total noise power. 

The signals on the $K$ available links are aggregated by applying maximum ratio combining~(MRC) to optimize signal reception and enhance diversity. 
The resulting SNR at ${m}$-th node after the MRC operation is denoted as $\gamma_{R_{m}}$ and is expressed by
    \begin{equation}\label{eq:SNRRAMO}
    \gamma_{R_m} = \frac{P_{\text{PLC}}\beta_{\text{PLC}}}{\sigma_{\text{PLC}}^2} \sum_{k=1}^{K}  h_{R_{m,k}}^2 = \bar{\gamma}_{R} \sum_{k=1}^{K}  h_{R_{m,k}}^2,
    \end{equation}
where \( \bar{\gamma}_{R} \triangleq \mathbb{E}\{\gamma_{R_m}\} = {P_{\text{PLC}}\beta_{\text{PLC}}}/{\sigma_{\text{PLC}}^2} \)  represents the average SNR per branch. 
Equation \eqref{eq:SNRRAMO} highlights that the total received SNR is obtained as a summation of the SNR contributions from all \( K \) nodes. 
Considering the form of the SNR expressed in~\eqref{eq:SNRRAMO} and knowing that the square of a lognormal RV is also a lognormal RV, we leverage the lognormal sum distribution approximation presented in~\cite[Eq.~(11)]{Beaulieu2004} to statistically characterize $\gamma_{R_{m}}$. 
This approach enables us to determine the PDF and CDF of \( \gamma_{R_m} \) in closed-form expressions accurately. 

Considering the $M$ signals, corresponding to the SNRs $\gamma_{R_{m}}$, selecting the maximum SNR among the $M$ available nodes is realized to optimize the system performance. 
The resulting cascaded PLC system SNR is then computed as
    \begin{equation}
        \gamma_{\text{PLC}} = \max\limits_{m \in \{1,\dots, M\}}( \gamma_{R_{m}} ).
    \end{equation}
Following this selection strategy, the CDF of $\gamma_{\text{PLC}}$ is given by~\cite[Eq. (6)]{Ai2021}
    \begin{align}\label{eq:CDFPLC}
    F_{\Gamma_{\text{PLC}}}(\gamma)  &= \left[ \Phi\left(a_0 -a_1\left(\gamma/\bar{\gamma}_{R}\right)^{-\frac{a_2}{\kappa}}\right)\right]^{M},
    \end{align}
where \( \Phi(\cdot) \) represents the CDF of a standard Gaussian RV with zero mean and unit variance~\cite[Eq. (8.250.1)]{Gradshteyn2007}, $\kappa = \ln{(10)}/10$, and the constants $a_{0}$, $a_{1}$, and $a_{2}$ are derived from the lognormal sum approximation, as detailed in~\cite{Beaulieu2004}. 
Similarly, the PDF of the PLC system equivalent to the SNR is expressed as~\cite[Eq. (7)]{Ai2021}
    \begin{align}\label{eq:PDFPLC}
    f_{\Gamma_{\text{PLC}}}(\gamma)  &= \frac{Ma_1a_2}{\kappa\sqrt{2\pi}\bar{\gamma}_{R}^{-\frac{a_2}{\kappa}}}\gamma^{-\left(\frac{a_2}{\kappa}+1\right)}e^{-\frac{1}{2}\left[a_0-a_1\left(\gamma/\bar{\gamma}_{R}\right)^{-\frac{a_2}{\kappa}}\right]^2}\nonumber\\
    &\times \left[ \Phi\left(a_0 -a_1\left(\gamma/\bar{\gamma}_{R}\right)^{-\frac{a_2}{\kappa}}\right)\right]^{M-1}.
    \end{align}  

\subsection{Cascaded Multiple-VLC Channel}

Within the framework of VLC systems, geometric and channel parameters significantly impact the performance of a wireless optical link. 
Consider a scenario where an LED access point communicates with a mobile user, as illustrated in the geometry presented in Fig.~\ref{fig:cascadedSystem}. 
The angles of irradiance ($\phi_u$) and incidence ($\psi_u$) at the user’s location, denoted as the $u$-th user, represent the angles at which light is emitted from the LED and received by the user’s photodetector (PD), respectively. 
Relative to the reference polar axis, the user is positioned at a horizontal distance $r_{u}$, forming an angle $\theta_{u}$  with the axis. 
The vertical separation between the LED and the user is denoted as $L$. 
Consequently, the optical link Euclidean distance is given by $d_{u} = \sqrt{L^2 + r_{u}^2}$. 
The LED’s coverage area is a circular cell with a radius \(r_\text{e}\), defining the illumination boundary.

The SNR for the direct current (DC) component of the LoS for the VLC link is expressed as
$\gamma_{\text{VLC}} = \bar{\gamma}_{\text{VLC}}  h_{u}^2,$
where \( \bar{\gamma}_{\text{VLC}} = {P_{\text{VLC}} \rho}/{\sigma_{\text{VLC}}^2} \) represents the average SNR per bit. 
Here, $P_{\text{VLC}}$ is the LED’s transmitted power, $\rho$ is the electrical-to-optical conversion efficiency, and $\sigma_{\text{VLC}}^2$ is the noise power at the end user. 
The DC channel gain $h_u$ quantifies the received optical power and is modeled as $   h_u = \frac{\Xi(m_\text{v} + 1)L^{m_\text{v} + 1}}{(r_u^2 + L^2)^{\frac{m_\text{v} + 3}{2}}},$
where \( \Xi = A R_p U(\psi_u) g(\psi_u)/(2\pi) \). 
The term $\Xi$ consolidates the PD’s physical area ($A$), responsivity ($R_p$), and gains from the optical filter \(U(\psi_u)\) and concentrator \(g(\psi_u)\). 
The gain of the concentrator is formulated as 
\begin{equation}
    \label{eq:conc_gain}
    g(\psi_{u}) = \begin{cases}
        \eta^2 / \sin^2( \Psi ),& \text{if\;} 0 \leq \psi_{u} \leq \Psi, \\
        0, & \text{if\;} \psi_{u} > \Psi,
    \end{cases}
\end{equation}
in which $\Psi$ is the field-of-view (FOV) angle of the PD and $\eta$ is the reffractive index of the medium. 
The parameter $m_\text{v}$, known as the Lambertian order, characterizes the LED’s radiation pattern and is determined by its semiangle $\phi_{1/2}$, which is the angle at which luminous intensity halves, as $m_\text{v} = -\frac{1}{\log_2(\cos \phi_{1/2})}.$
A smaller value of $\phi_{1/2}$ corresponds to a more directional LED beam, i.e. higher $m_\text{v}$. 

Assuming users are uniformly distributed within the LED’s coverage area, the horizontal distance $r_u$ becomes an RV with PDF defined as~\cite{Gupta2017}
    \begin{equation}
    f_{r_u}(r) = \frac{2r}{r_\text{e}^2}, \quad 0 \leq r \leq r_\text{e}. 
    \end{equation} 
By substituting $r_u$ into the SNR and channel gain equations, the PDF of the SNR $\gamma_{\text{VLC}}$ is derived using a RV transformation. 
The resulting PDF is a power-law function given by~\cite[Eq. (5)]{Gupta2017}
    \begin{equation}
    \label{eq:vlc_pdf}
    f_{\Gamma_{\text{VLC}}}(\gamma) = \frac{[\Xi(m_\text{v} + 1)L^{m_\text{v} + 1}\sqrt{\bar{\gamma}_{\text{VLC}}}]^{\frac{2}{m_\text{v} + 3}}}{r_\text{e}^2(m_\text{v} + 3)} \gamma^{-\frac{m_\text{v} + 4}{m_\text{v} + 3}}, 
    \end{equation}
for $\gamma_{\text{e}} \leq \gamma \leq \gamma_{\text{c}}$, in which $\gamma_{\text{e}}$ is the minimum SNR for users at the cell edge, i.e. $r_{u} = r_\text{e}$, and $\gamma_{\text{c}}$ is the maximum SNR when users are directly under the LED, i.e. $r_{u} = 0$, at the cell center. 
These bounds are quantified by
    \begin{equation}
    \gamma_{\text{e}} = \frac{\bar{\gamma}_{\text{VLC}}[\Xi(m_\text{v} + 1)L^{m_\text{v} + 1}]^2}{(r_\text{e}^2 + L^2)^{m_\text{v} + 3}}
    \end{equation}
and
    \begin{equation}
    \gamma_{\text{c}} = \frac{\bar{\gamma}_{\text{VLC}}[\Xi(m_\text{v} + 1)L^{m_\text{v} + 1}]^2}{L^{2(m_\text{v} + 3)}}.
    \end{equation}
For $\gamma < \gamma_{\text{e}}$ and $\gamma > \gamma_{\text{c}}$, $f_{\Gamma_{\text{VLC}}}(\gamma) = 0$.

The CDF of $\gamma_{\text{VLC}}$ is found by integrating the PDF expressed in~\eqref{eq:vlc_pdf}, which results in
    \begin{equation}
    \label{eq:vlc_cdf}
    F_{\Gamma_{\text{VLC}}}(\gamma) =  \varepsilon - \upsilon\left(\frac{\gamma}{\bar{\gamma}_{\rm VLC}}\right)^{-\frac{1}{m_\text{v}+3}},
    \end{equation}
for $\gamma_{\text{e}} \leq \gamma \leq \gamma_{\text{c}}$, where \( \varepsilon = {(r_\text{e}^2 + L^2)}/{r_\text{e}^2} \) and \( \upsilon = {[\Xi(m_\text{v}+ 1)L^{m_\text{v} + 1}]^{\frac{2}{m_\text{v} + 3}}}/{r_\text{e}^2} \). 
These parameters encapsulate the geometric and channel characteristics, enabling a compact representation of the SNR distribution for the VLC link. 
Due to the SNR constraints, limited to the range $[\gamma_{\text{e}},\gamma_{\text{c}}]$, $F_{\Gamma_{\text{VLC}}}(\gamma) = 0$ for $\gamma < \gamma_{\text{e}}$ and $F_{\Gamma_{\text{VLC}}}(\gamma) = 1$ for $\gamma > \gamma_{\text{c}}$.

Considering that the indoor system contains $N$ LEDs, corresponding to $N$ SNR terms, a given user connects to the LED that maximizes the SNR among the $N$ links, corresponding to the closest access point according to the adopted model. 
Assuming that the $N$ links promote i.i.d. SNRs terms, characterized by~\eqref{eq:vlc_pdf} and~\eqref{eq:vlc_cdf}, the CDF and PDF of SNR experienced by the user, denoted as $\gamma_{\text{VLC}}^{\text{MAX}}$, are derived by considering the joint statistics of all links.
Using binomial expansion, these are  
\begin{equation}\label{eq:CDFVLC}
F_{\Gamma_{\text{VLC}}^{\text{MAX}}}(\gamma) = \sum_{i=0}^{N} (-1)^i \binom{N}{i} \varepsilon^{N-i} \upsilon^i \left(\frac{\gamma}{\bar{\gamma}_{\text{VLC}}}\right)^{-\frac{i}{m_\text{v} + 3}}
\end{equation}
and
\begin{equation}\label{eq:pDFVLC}
f_{\Gamma_{\text{VLC}}^{\text{MAX}}}(\gamma) = \frac{N}{m_\text{v} + 3} \sum_{i=0}^{N-1} (-1)^i \binom{N-1}{i} \frac{\varepsilon^{N-i-1} \upsilon^{i+1}}{\bar{\gamma}_{\text{VLC}}^{-\frac{i + 1}{m_\text{v} + 3}} \gamma^{\frac{m_\text{v} + 4 + i}{m_\text{v} + 3}}}.
\end{equation}

Both expressions in~\eqref{eq:CDFVLC} and~\eqref{eq:pDFVLC} are valid within the interval $[\gamma_{\text{e}},\gamma_{\text{c}}]$.
In turn, for $\gamma<\gamma_{\text{e}}$, $F_{\Gamma_{\text{VLC}}^{\text{MAX}}}(\gamma)=0$ and $f_{\Gamma_{\text{VLC}}^{\text{MAX}}}(\gamma)=0$ and for $\gamma>\gamma_{\text{e}}$, $F_{\Gamma_{\text{VLC}}^{\text{MAX}}}(\gamma)=1$ and $f_{\Gamma_{\text{VLC}}^{\text{MAX}}}(\gamma)=0$.
\section{End-to-End SNR Statistics}\label{Statistics}

The PLC and VLC systems are intermediated by a DF relay, promoting a resulting equivalent SNR equal to $\Gamma_{\text{eq}} = {\min}(\Gamma_{\rm PLC}, \Gamma_{\rm VLC}^{\rm MAX})$, that is, the minimum between the output SNRs of the individual systems. 
Assuming that $\gamma_{\text{PLC}}$ and $\gamma_{\text{VLC}}^{\text{MAX}}$ are independent, the CDF of the equivalent end-to-end SNR is generally given by $      F_{\Gamma_{\text{eq}}}(\gamma)  = F_{\Gamma_{\text{PLC}}}(\gamma) + F_{\Gamma_{\text{VLC}}^{\text{MAX}}}(\gamma) - F_{\Gamma_{\text{PLC}}}(\gamma)F_{\Gamma_{\text{VLC}}^{\text{MAX}}}(\gamma).$
Using \eqref{eq:CDFPLC} and \eqref{eq:CDFVLC}, it follows that
\begin{align}\label{eq:CDFENDTOEND}
    F_{\Gamma_{\text{eq}}} (\gamma)  &= \left[ \Phi\left(a_0 -a_1\left(\gamma/\bar{\gamma}_{R}\right)^{-\frac{a_2}{\kappa}}\right)\right]^{M}\nonumber\\
    &+ \sum_{i=0}^{N}(-1)^i\binom{N}{i}\varepsilon^{N-i} \upsilon^i\left(\gamma/\bar{\gamma}_{\text{VLC}}\right)^{-\frac{i}{m_\text{v} + 3}}\nonumber\\
    &-\left[ \Phi\left(a_0 -a_1\left(\gamma/\bar{\gamma}_{R}\right)^{-\frac{a_2}{\kappa}}\right)\right]^{M}\nonumber\\
&\times\sum_{i=0}^{N}(-1)^i\binom{N}{i}\varepsilon^{N-i} \upsilon^i\left(\gamma/\bar{\gamma}_{\text{VLC}}\right)^{-\frac{i}{m_\text{v} + 3}},
\end{align}
in the interval $\gamma_{\text{e}} \leq \gamma \leq \gamma_{\text{c}}$.

Differentiating $F_{\Gamma_{\text{eq}}}(\gamma)$ with respect to $\gamma$, the PDF $f_{\Gamma_{\text{eq}}} (\gamma)$ can be written as
    \begin{align}\label{eq:prePDF}
      f_{\Gamma_{\text{eq}}}(\gamma)  &= f_{\Gamma_{\text{PLC}}}(\gamma) + f_{\Gamma_{\text{VLC}}^{\text{MAX}}}(\gamma)\nonumber\\
      &- f_{\Gamma_{\text{PLC}}}(\gamma)F_{\Gamma_{\text{VLC}}^{\text{MAX}}}(\gamma) - F_{\Gamma_{\text{PLC}}}(\gamma)f_{\Gamma_{\text{VLC}}^{\text{MAX}}}(\gamma).
    \end{align}
Replacing~(\ref{eq:CDFPLC}),~(\ref{eq:PDFPLC}),~(\ref{eq:CDFVLC}), and~(\ref{eq:pDFVLC}) in~(\ref{eq:prePDF}), (\ref{eq:PDFENDTOEND}) can be obtained.

\begin{figure*}
 \begin{align}\label{eq:PDFENDTOEND}
    f_{\Gamma_{\text{eq}}} (\gamma)  &= \frac{Ma_1a_2}{\kappa\sqrt{2\pi}\bar{\gamma}_{R}}\frac{\left(\gamma/\bar{\gamma}_{R}\right)^{-\left(\frac{a_2}{\kappa}+1\right)}}{e^{\frac{1}{2}\left[a_0-a_1\left(\gamma/\bar{\gamma}_{R}\right)^{-\frac{a_2}{\kappa}}\right]^2}}\left[ \Phi\left(a_0 -a_1\left(\gamma/\bar{\gamma}_{R}\right)^{-\frac{a_2}{\kappa}}\right)\right]^{M-1}\left( 1 -  \sum_{i=0}^{N}(-1)^i\binom{N}{i}\varepsilon^{N-i} \upsilon^i\left(\gamma/\bar{\gamma}_{\rm VLC}\right)^{-\frac{i}{m_\text{v} + 3}}        \right) \nonumber\\
    &+ \frac{N}{m_\text{v}+3}\sum_{i=0}^{N-1}(-1)^i\binom{N-1}{i}\varepsilon^{N-i-1} \upsilon^{i+1}\gamma^{-\frac{m_\text{v}+4+i}{m_\text{v}+3}}\bar{\gamma}_\text{VLC}^{\frac{i+1}{m_\text{v}+3}}\left(1 - \left[ \Phi\left(a_0 -a_1\left(\gamma/\bar{\gamma}_{R}\right)^{-\frac{a_2}{\kappa}}\right)\right]^{M} \right).
\end{align}
   \hrulefill
\end{figure*}

\section{Performance Metrics}\label{PerformanceAnalysis}

\subsection{Outage Probability}
The OP, considering the proposed system, is denoted by $P_{\text{out}} = F_{\Gamma_{\rm eq}}(\gamma_{\text{th}})$ and can be calculated from~(\ref{eq:CDFENDTOEND}) as
\begin{align}
    P_{\text{out}}  &= \left[ \Phi\left(a_0 -a_1\left(\gamma_{\text{th}}/\bar{\gamma}_{R}\right)^{-\frac{a_2}{\kappa}}\right)\right]^{M}\nonumber\\
    &+ \sum_{i=0}^{N}(-1)^i\binom{N}{i}\varepsilon^{N-i} \upsilon^i\left(\gamma_{\text{th}}/\bar{\gamma}_{\text{VLC}}\right)^{-\frac{i}{m_\text{v} + 3}}\nonumber\\
    &-\left[ \Phi\left(a_0 -a_1\left(\gamma_{\text{th}}/\bar{\gamma}_{R}\right)^{-\frac{a_2}{\kappa}}\right)\right]^{M}\nonumber\\
    &\times\sum_{i=0}^{N}(-1)^i\binom{N}{i}\varepsilon^{N-i} \upsilon^i\left(\gamma_{\text{th}}/\bar{\gamma}_{\text{VLC}}\right)^{-\frac{i}{m_\text{v} + 3}},
    \end{align}
in which $\gamma_{\text{th}}$ is a specified threshold.

\subsection{Average BEP}

The average BEP can be calculated for binary constellations as~\cite[Eq. (12)]{Ansari}
    \begin{align}\label{eq:PeCDF}
        P_{\text{e}} = \frac{q^p}{2\Gamma(p)}\int_{0}^{\infty}\frac{e^{-q\gamma}}{\gamma^{1-p}}F_{\Gamma_{\text{eq}}}(\gamma)\text{d}\gamma,
    \end{align}
in which $\Gamma(\cdot)$ denotes the Gamma function and the parameters $p$ and $q$ are determined by the specific modulation scheme. 
For instance, $p=0.5$ and $q=1$ for a binary phase shift keying (BPSK) constellation~\cite{Ansari}.

Replacing~(\ref{eq:CDFENDTOEND}) in~(\ref{eq:PeCDF}) and considering the intervals of the function $F_{\Gamma_{\text{VLC}}^{\text{MAX}}}(\gamma)$, it follows that $  P_{\text{e}} = \frac{\mathcal{I}_{1} - \mathcal{I}_{2} +\mathcal{I}_{3}+\mathcal{I}_{4}}{2q^{-p}\Gamma(p)},$
with 
    \begin{align}\label{eq:I1}
        \mathcal{I}_{1} = \int_{0}^{\gamma_{\text{c}}} \gamma^{p-1} e^{-q\gamma}
        \left[ \Phi\left(a_0 - a_1 \left( \frac{\gamma}{\bar{\gamma}_{R}} \right)^{-\frac{a_2}{\kappa}}\right)\right]^M \text{d}\gamma,
    \end{align}
    \begin{align}\label{eq:I2}
        \mathcal{I}_{2} &= \sum_{i=0}^{N} (-1)^i \binom{N}{i} \varepsilon^{N-i} \upsilon^i 
        \left( \frac{1}{\bar\gamma_{\text{VLC}}} \right)^{-\frac{i}{m_\text{v} + 3}}\nonumber\\
        &\times     \int_{\gamma_{\text{e}}}^{\gamma_{\text{c}}} \frac{\gamma^{p-1-\frac{i}{m_\text{v} + 3}}}{e^{q\gamma}}
        \left[ \Phi\left(a_0 - a_1 \left( \frac{\gamma}{\bar{\gamma}_{R}} \right)^{-\frac{a_2}{\kappa}}\right)\right]^M\text{d}\gamma,
    \end{align}
    \begin{align}\label{eq:I3}
        \mathcal{I}_{3} &=  \sum_{i=0}^{N} (-1)^i \binom{N}{i} \varepsilon^{N-i} \upsilon^i 
        \left( \frac{1}{\bar\gamma_{\text{VLC}}} \right)^{-\frac{i}{m_\text{v} + 3}}\nonumber\\
        &\times \int_{\gamma_{\text{e}}}^{\gamma_{\text{c}}} \gamma^{p-1-\frac{i}{m_\text{v} + 3}} e^{-q\gamma} \text{d}\gamma,
    \end{align}
and $\mathcal{I}_{4} = \int_{\gamma_{\text{c}}}^{\infty} \gamma^{p-1} e^{-q\gamma} \text{d}\gamma.$

Changing the integration intervals to $[-1,1]$ by means of~\cite[Eq. (3.021)]{Gradshteyn2007}, and using the $N_{a}$-order Gauss-Legendre quadrature with roots $x_{j}$ and weights $w_{j}$, it follows that
$\mathcal{I}_1  = \frac{\gamma_{\text{c}}}{2}\tilde{\mathcal{I}}_{1}\left(\frac{\gamma_{\text{c}}}{2}x_j+\frac{\gamma_{\text{c}}}{2}\right),$

\begin{align}
    \mathcal{I}_2  &= \sum_{i=0}^{N} (-1)^i \binom{N}{i} \varepsilon^{N-i} \upsilon^i 
    \left( \frac{1}{\bar\gamma_{\text{VLC}}} \right)^{-\frac{i}{m_\text{v} + 3}} \frac{(\gamma_{\text{c}}-\gamma_{\text{e}})}{2}\nonumber\\
    \end{align}
and
    \begin{align}
        \mathcal{I}_3  &= \sum_{i=0}^{N} (-1)^i \binom{N}{i} \varepsilon^{N-i} \upsilon^i 
        \left( \frac{1}{\bar\gamma_{\text{VLC}}} \right)^{-\frac{i}{m_\text{v} + 3}} \frac{(\gamma_{\text{c}}-\gamma_{\text{e}})}{2}\nonumber\\
    &\times\tilde{\mathcal{I}}_{3}\left(\frac{\gamma_{\text{c}}-\gamma_{\text{e}}}{2}x_j+\frac{\gamma_{\text{c}}+\gamma_{\text{e}}}{2}\right),
        \end{align}
in which
\begin{align}
    \tilde{\mathcal{I}}_{1}(x) &= \sum_{j=1}^{N_a} w_j x^{p-1}e^{-qx} \left[ \Phi\left(a_0 -a_1\left(x/\bar{\gamma}_{R}\right)^{-\frac{a_2}{\kappa}}\right)\right]^{M},
\end{align}
\begin{align}
    \tilde{\mathcal{I}}_{2}(x) &= \sum_{j=1}^{N_a} w_j x^{p-1-\frac{i}{m_\text{v}+3}}e^{-qx} \left[ \Phi\left(a_0 -a_1\left(x/\bar{\gamma}_{R}\right)^{-\frac{a_2}{\kappa}}\right)\right]^{M},
\end{align}
and $ \tilde{\mathcal{I}}_{3}(x) = \sum_{j=1}^{N_a} w_j x^{p-1-\frac{i}{m_\text{v}+3}}e^{-qx}.$
In turn, by means of~\cite[Eq. (8.350.2)]{Gradshteyn2007},
$\mathcal{I}_4    =\frac{1}{q^{p}}\Gamma(p, q\gamma_{\text{c}}),$
in which $\Gamma(\cdot, \cdot)$ is the upper incomplete Gamma function.


\section{Results}\label{results}

This section presents the numerical results regarding the performance metrics of the proposed system. 
The validity of the derived analytical expressions is confirmed through Monte Carlo simulations developed in Python\footnote {The code used for the simulations is publicly available at: \url{https://github.com/HigoTh/PLC_VLC}}.
The presented analytical curves closely align with the simulation results.
The values of the system parameters adopted here in the results, and also elsewhere in the literature \cite{Ai2021,Ndjiongue,Jani2019,Jani2020,Jani2021}, are summarized in Table~\ref{tab:1}.

\begin{table}
\centering
\caption{Simulation Parameters.}
\label{tab:1}
\begin{tabular}{cc}
\hline
\textbf{Parameter} & \textbf{Values}                            \\ 
\hline
\hline
Powerline attenuation parameters -- $(\alpha_{1},\alpha_{2})$ & ($0.0093$, $0.0051$) \\ \hline
PLC channels statistics -- $(\mu_{h}, \sigma_{h}^2)$ & ($-1.549$, $1$) \\ \hline
Impulsive noise probability -- $p$ & $0.05$ \\ \hline
Powerline length -- $\ell_{\text{PLC}}$ & $5$~m \\ \hline
PLC operation frequency -- $f$ & $20$ MHz \\ \hline
PD physical area -- $A$ & $10^{-4}$~{m}$^2$ \\ \hline
Optical filter gain -- $U(\psi_{u})$ & $1$ \\ \hline
Reffractive index -- $\eta$ & $1.5$ \\ \hline
Electrical-to-optical conversion efficiency -- $\rho$ & $0.64$ A/W \\ \hline
Cell radius -- $r_\text{e}$  & $2.5$~{m} \\ \hline
\hline
\end{tabular}
\end{table}

\begin{figure}
\centering
\includegraphics[width=0.87\linewidth]{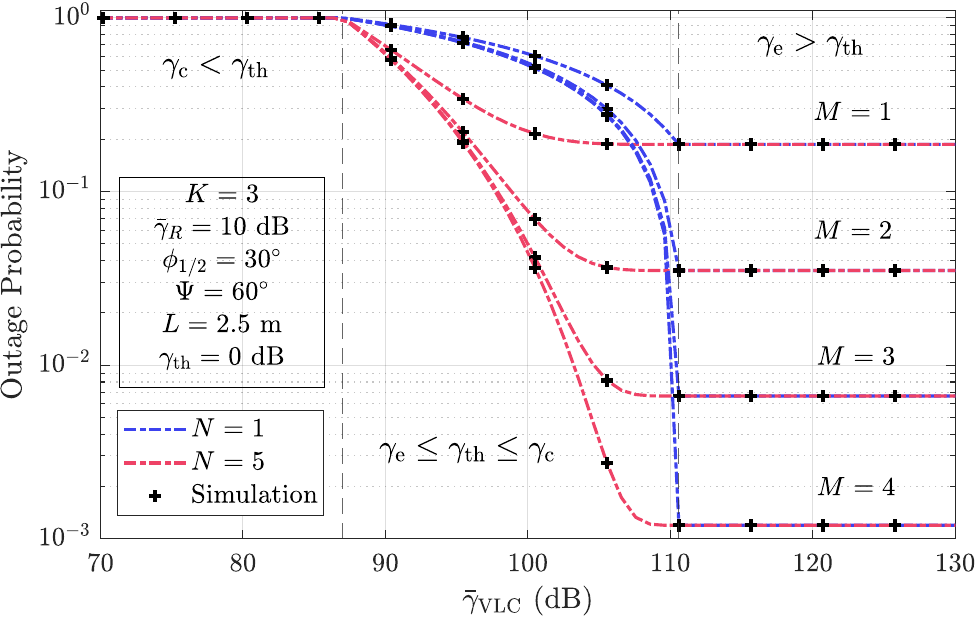}
\caption{OP curves as a function of $\bar{\gamma}_{\text{VLC}}$ for different configurations of the number of branches $M$ and the number of LEDs $N$.}
\label{fig:fig_1}
\end{figure}

Fig.~\ref{fig:fig_1} illustrates the OP curves as a function of $\bar{\gamma}_{\text{VLC}}$ for different configurations of the number of branches $M$ on the PLC subsystem and the number of LEDs $N$ on the VLC side. 
The remaining system parameters are set as follows: $K = 3$, $\bar{\gamma}_R = 10$~dB, $\phi_{1/2} = 30^{\circ}$, $\Psi = 60^{\circ}$, and $\gamma_\text{th} = 0$~dB. 
The OP curves exhibit three distinct regions, a behavior attributable to the use of a DF relay between the PLC and VLC systems, as well as the constraint that the SNR of the VLC is bounded within the interval $[\gamma_{\text{e}}, \gamma_{\text{c}}]$. 
The equivalent SNR in the first region is dominated by the VLC channel, i.e., $\gamma_{\text{eq}} \approx \gamma_{\text{VLC}}^{\text{MAX}}$, since the SNR of the VLC is significantly lower than that of the PLC. 
In this region, the upper bound $\gamma_{\text{c}}$ is below the target threshold $\gamma_\text{th}$, leading to a near-certain outage.

As $\bar{\gamma}_{\text{VLC}}$ increases, $\gamma_{\text{c}}$ surpasses $\gamma_\text{th}$, and the SNRs of both the PLC and VLC links become comparable. 
This results in a gradual reduction in the OP, reflecting the improved reliability of the combined system in the second region. 
At high values of $\bar{\gamma}_{\text{VLC}}$, the SNR of the VLC becomes significantly greater than that of the PLC. 
In this regime, the system performance is limited by the PLC link, and thus the equivalent SNR converges to $\gamma_{\text{eq}} \approx \gamma_{\text{PLC}}$. 
Consequently, further increases in $\bar{\gamma}_{\text{VLC}}$ have a negligible impact on the OP. 

The impact of the number of branches in the system is also evident in Fig.~\ref{fig:fig_1}.
Specifically, increasing the number of branches $M$ on the PLC side leads to a reduction in the OP, as the system benefits from greater spatial diversity. 
Similarly, an increase in the number of LEDs $N$ on the VLC side also results in lower OP values, owing to enhanced optical diversity at the receiver. 
This result illustrates how specific configurations of the proposed system influence performance, highlighting the benefits of employing multiple branches in the PLC system and multiple LEDs in the VLC system, and confirms that the proposed system architecture generalizes and extends different scenarios of the cascaded PLC/VLC system, offering significant performance gains.

\begin{figure}
    \centering
    \includegraphics[width=0.87\linewidth]{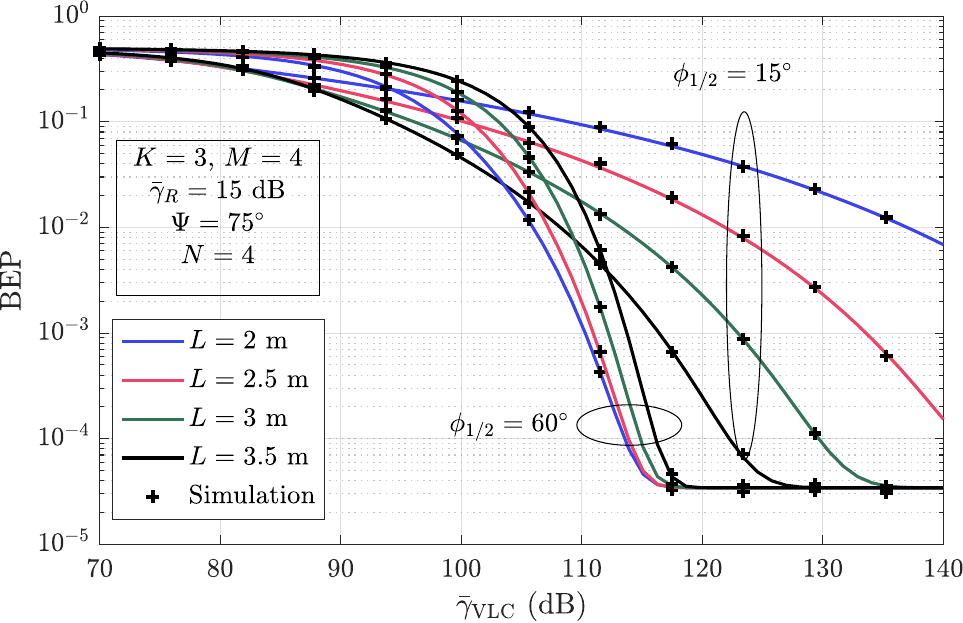}
    \caption{Average BEP curves as a function of $\bar{\gamma}_{\text{VLC}}$ for different values of the semiangle $\phi_{1/2}$ and vertical length $L$.}
    \label{fig:fig_1b}
\end{figure}

Fig.~\ref{fig:fig_1b} presents average BEP curves as a function of $\bar{\gamma}_{\text{VLC}}$, for different values of $\phi_{1/2}$ and $L$, considering $K = 3$, $\bar{\gamma}_R = 15$~dB, $\Psi = 75^{\circ}$, $M=4$, and $N = 4$. 
In the case of a wide angular opening, with $\phi_{1/2}=60^{\circ}$, increasing the vertical distance $L$ degrades the system performance, resulting in higher BEP values. 
For this configuration, the $60^{\circ}$ beamwidth provides full cell coverage ($r_\text{e}=2.5$~m) without significant visibility loss, keeping the entire cell within the LED's half-power zone for the considered $L$ values. 
Conversely, for a narrower beamwidth ($\phi_{1/2}=15^{\circ}$), the half-power coverage area decreases at shorter distances, thereby increasing the probability that users will be outside this zone. 
In this scenario, increasing $L$ expands the LED's visibility area, consequently reducing the observed BEP and improving performance. 

Note that this behavior is valid only for the range of $L$ values considered in the analysis. 
However, these observations have limited validity across all possible values of $L$ and cell sizes $r_\text{e}$. 
For the narrow beam case ($\phi_{1/2}=15^{\circ}$), there exists a critical distance beyond which the entire cell $r_\text{e}$ falls within the LED's half-power angle. 
Beyond this threshold distance, the BEP improvement trend would no longer hold, as the system would instead begin to exhibit the same coverage-limited behavior observed in the wide beam case.

\section{Conclusions}\label{conclusao}

This paper has presented a cascaded multiwire-PLC/multiple-VLC system for indoor environments.
By leveraging the widespread availability of electrical wiring and the dual-use functionality of light-emitting diodes, the proposed system offers enhanced performance and a cost-effective solution for data communication coverage and reliability.
New analytical expressions have also been derived for end-to-end signal-to-noise ratio statistics, as well as for the OP and average BEP metrics. 
Monte Carlo simulations confirmed the theoretical results.



\balance

\end{document}